\documentclass[12pt,preprint]{aastex}
\usepackage{amsmath}
\usepackage{natbib}
\usepackage{amssymb}
\begin{document}
\title{Standardization, Distance, Host Galaxy Extinction of Type Ia Supernova and Hubble Diagram from the Flux Ratio Method} 
%\author{}

\author{Bo Yu\altaffilmark{1,2,3}, Gui-Lin Yang\altaffilmark{2,3} and Tan Lu\altaffilmark{1,2,3}}
\altaffiltext{1} {Purple Mountain Observatory, Chinese Academy of Sciences,
Nanjing 210008, China}
\altaffiltext{2} {Department of Physics, Nanjing University, Nanjing 210093, China}
\altaffiltext{3} {Joint Center for Particle, Nuclear Physics and Cosmology,
Nanjing University-Purple Mountain Observatory, Nanjing 210093, China}

\begin{abstract}
In this paper we generalize the flux ratio method \cite{2009A&A...500L..17B} to the case of
two luminosity indicators and search the optimal luminosity-flux ratio relations on
a set of spectra whose phases are around not only the date of bright light but
also other time. With these relations, a new method is proposed to constrain
the host galaxy extinction of SN Ia and its distance. It is first applied to the
low redshift supernovas and then to the high redshift ones. The results of the
low redshift supernovas indicate that the flux ratio method can indeed give
well constraint on the host galaxy extinction parameter E(B-V), but weaker constraints on
$R_{V}$. The high redshift supernova spectra are processed by the same method as the
low redshift ones besides some differences due to their high redshift. Among 16
high redshift supernovas, 15 are fitted very well except 03D1gt. Based
on these distances, Hubble diagram is drew and the contents of the Universe are analyzed. It
supports an acceleration behavior in the late Universe. Therefore, the flux ratio
method can give constraints on the host galaxy extinction and
supernova distance independently. We believe, through further studies, it may provide a
precise tool to probe the acceleration of the Universe than before.

\end{abstract}


\section{Introduction}
\label{sn:introduction}
The Type Ia Supernova (SN Ia) as a standard candle plays an important role
in the cosmological probes. It is the relations proposed in 
\cite{1993ApJ...413L.105P} that make the measurement of decelerator
factor of the universe  possible. Based on this relationship,
a set of improved techniques are developed, such as the Multicolor
Light-Curve Shape method (MLCS) \cite{1996ApJ...473...88R}, the Stretch
method \cite{1997ApJ...483..565P}, the Spectral Adaptive Light curve
Template method (SALT) \cite{2005A&A...443..781G} and so on. 
Using these methods, the luminosity distance of a supernova can be
measured to a precision $7-10 \%$. In contrast to these methods which
use the light curve shape and color information to standardize the supernova
luminosity, some authors use spectral features as the luminosity indicators,
for example, \cite{1995ApJ...455L.147N}, \cite{2006ApJ...647..513B}, 
\cite{2006MNRAS.370..299H}, \cite{2008A&A...477..717B}. Although these
methods are very interesting, till now they are not so much competitive
as the above mentioned methods.

The SN Ia luminosity is affected by many factors. One of the most important is the host galaxy
 extinction. As for this, there exits several methods to estimate it. The first one is the
 Lira-Phillips relation \cite{1999AJ....118.1766P}, which show that the B-V color curve of
 SN Ia with small host galaxy extinction evolves in a similar way within [30-90] days after
 the maximum light. Therefore, comparing the observed B-V color curve with this color
 evolution, one can constrain the host galaxy extinction for an individual SN. The second
 one is the Multicolor Light Curves Shape (MLCS) method \cite{1996ApJ...473...88R}, which
 compare the observed light curves with the template light curves. The third one is to
 measure the equivalent width of the interstellar Na I D doublet which is correlate to
 the line of sight dust \cite{1990A&A...237...79B,1997A&A...318..269M}.

Recently, \cite{2009A&A...500L..17B} searched the optimal correlation between the B maximum
 M and the flux ratios at any two wavelengths. It is claimed that the scatter can be decreased
 to 0.12. However, in that paper, the phases of spectra are required within $\pm 2.5$ days
 around the day of B maximum $t_0$, and the host galaxy dust extinction are corrected in an indirect method. Here
 we search the possible relationships between the B maximum M and the flux ratios at different
 phases, for example, $t = -6 \sim 14$ days, with all the supernova spectra are deredden to
 correct for both Galaxy and host galaxy extinction directly. Furthermore, we search the optimal linear
 relationships by introducing two flux ratios $R_{\lambda_1 / \lambda_2},R_{\lambda_3 / \lambda_4}$
 as the luminosity indicators. It is expected that a part of the intrinsic scatter in the one
 luminosity indicator case can be compensated by introducing another luminosity indicator,
 therefore the final scatter can be deceased as much as possible.

After corrected for the host galaxy dust contamination directly, small luminosity scatters of SN
 Ia can be obtained from the flux ratio method, especially in the case of two flux ratios
 being used as the luminosity indicators. With this, it is possible to fit the host galaxy
 extinction and distance modulus at the same time. Such an attempt to constrain the host
 galaxy extinction and distance modulus is different from the methods employed before. It
 can provide another independent estimation on host galaxy extinction and distance modulus.
 Besides this, several fitting results for the same supernova can be compared with each other
 since the luminosity-flux ratio (L-F) relation has been generalized to other supernova phases
 besides the maximum day.

The aim of this paper is to develop the flux ratio method and make it to be a
reliable tool to estimate the distance of type Ia supernova and its host
galaxy extinction. We will go through all the procedures to produce a
Hubble diagram from the supernova spectra by using the flux ratio method.
Considering that the number of the train spectra is finite and it is the first time to  
apply the method, we pay more attentions on the method itself than the
detailed results.  

This paper is organized as follows. Section \ref{sn:relation} concentrates on the
luminosity-flux ratio relation, in which the SN Ia luminosities are fitted
with one luminosity indicator or two luminosity indicators respectively and the optimal
luminosity scatters are obtained by searching all the wavelength pairs at each day
within [-6,14]. In section \ref{sn:fitting}, a method is proposed to constrain
the host galaxy extinction and distance of the supernova. Then it is
applied to the low redshift samples. In section \ref{sn:highz}, the flux ratio method is
applied to the high redshift supernova spectra, and the difference between high redshift
supernova samples and the low ones are emphasized. At the end of this section, the Hubble
diagram and the $(\Omega_{m}, \Omega_{\Lambda})$ contour map are given. Finally we give
the discussion and summary in the last section.

\section{Luminosity-flux ratio relation}
\label{sn:relation}
\subsection{Data set}
\label{sn:dataset}
The supernova spectra used in this section come from two sources, one is
 \cite{2008AJ....135.1598M} and the other is the Online Supernova Spectrum
 Archive \footnote{http://bruford.nhn.ou.edu/~suspect/index1.html}. All of
 them are public available.

The supernovas used in this section are required to satisfy the following
conditions: 1) There are photometric observations within the time interval [-6,6] day
(relative to the date of B maximum here and after) and spectrographic observation
in the time interval [-6,14] day; 2) $A_{V,host} < 1.0$; 3) the difference
between the B magnitude synthesized from the supernova spectrum and the observed
B magnitude must be smaller than 0.05 mag. Totally, we obtained 38 supernovas which are
listed in Table \ref{tbl1}. In the same table, the following information is also listed:
the redshift $z_{cmb}$ (in the CMB rest frame), the phase windows each supernova contributed
to, the date of B maximum $t_{0}$, the apparent B maximum magnitude m, Milky Way dust
extinction $E(B-V)_{gal}$, and the host galaxy extinction parameters $(A_V,R_V)$. 
The Milky Way dust extinction parameter $E(B-V)_{gal}$ is adopted from the \cite{1998ApJ...500..525S},
while the host galaxy extinction parameter and the time $t_{0}$ comes mainly from the
results of \cite{2007ApJ...659..122J}, and for those supernovas which do not
appear in that paper these values are from the results fitted by individual authors.
Table \ref{tbl1x} is the information of the left supernovas which are not included in the
train data set since they do not satisfy the above condition 1) or 2) but can be used in later section.

The absolute magnitude M is calculated by
\begin{eqnarray}
\label{eq1}
M=m-k-A_{B,gal}-A_{B,host}-\mu,
\end{eqnarray}
where m is the apparent B band maximum magnitude, $k$ is the $K$ correction term
in the date $t_0$ which is calculated by the software SNANA \cite{2009PASP..121.1028K},
$\mu$ is the distance modulus, and $A_{B,gal}$ $(A_{B,host})$ is the Milky Way (host galaxy)
extinction in $B$ band (strictly, Equation \ref{eq1} should be
$M_{B}=m_{X}-k_{XB}-A_{X,gal}-A_{B,host}-\mu$, for low redshift supernova, X is B filter band.
In later section, we use the simple form. When it is applied to high redshift samples, 
it means the strict form). For all supernovas, the distance modulus is adopted from the results
of \cite{2007ApJ...659..122J} if it is possible, otherwise it is directly calculated from $z_{cmb}$.
In all cases $H_{0}$ is fixed at $65 km/s$, different choice of $H_{0}$ does not affect the
later analysis but just changes the zero point of the luminosity-flux ratio relation by a constant.
So after here, when $H_{0}$ is needed, it takes a value of $65 km/s$.

All the train spectra are deredshifted to the supernova rest frame, and deredden by using the Galaxy
 and host galaxy extinction presented in Table \ref{tbl1} through the CCM extinction law
 \cite{1989ApJ...345..245C}. Like \cite{2009A&A...500L..17B}, the spectrum flux are binned
 in $c\varDelta \lambda / \lambda \sim 2000 km/s$ with
 $F(\lambda) = \int_{\lambda - \Delta \lambda /2}^{\lambda + \Delta \lambda /2}
 f(\lambda) d\lambda / \varDelta \lambda$ ($f(\lambda)$ is the flux after deredshifted and deredden).

\begin{deluxetable}{llllllllll}
\tabletypesize{\scriptsize}
\rotate
\tablecaption{Supernova and Host Galaxy Information.}
\tablewidth{0pt}
\tablehead{
\colhead{SN Ia}   &
\colhead{$cz_{cmb}$}   &
\colhead{$HJD_{B_{max}}$}   &
\colhead{phase window}   &
\colhead{$\text{m}^{max}_{B}$ \tablenotemark{a}}   &
\colhead{$\mu$}  &
\colhead{$E(B-V)_{gal}$\tablenotemark{b}}   &
\colhead{$A_{V,host}$}  &
\colhead{$R_{V,host}$}  &
\colhead{Ref} \\
& km/s & -240000(day) & \,\,\,\,\,\,\,\,\,\, day & \,\,\,\,\, mag & \,\,\,\,\,\,\,\,\,\, mag & \,\,\,\,\, mag & \,\,\,\,\,\,\,\, mag &  &
}
\startdata
1981B  & 2151 & 44670.95(0.73) &  -3 $\sim$ 1, 14            &  12.00(0.03) & 31.104(0.140) & 0.018 & 0.364(0.121) & 3.1  & 1,3,4,5,6  \\
1992A  & 1781 & 48640.63(0.19) &  -6 $\sim$ 9, 14            &  12.56(0.02) & 31.654(0.067) & 0.018 & 0.038(0.032) & 3.1  & 1,3,7,8  \\
1994D  & 928  & 49432.47(0.10) &-6$\sim$0,2$\sim$6,10$\sim$14&  11.84(0.03) & 31.187(0.067) & 0.022 & 0.109(0.047) & 3.1  & 1,3,9,10,11  \\
1996X  & 2333 & 50190.85(0.33) &  -3 $\sim$ 7                &  13.20(0.02) & 32.432(0.065) & 0.069 & 0.061(0.044) & 3.1  & 1,3,12,13 \\
1997br & 2386 & 50559.26(0.23) &  -6,-5,6 $\sim$ 10          &  14.05(0.03) & 32.248(0.104) & 0.113 & 0.804(0.112) & 3.03 & 1,3,7,14,15 \\
1997cn & 5092 & 50586.64(0.77) &  2 $\sim$ 6                 &  17.20(0.10) & 34.532(0.070) & 0.027 & 0.071(0.059) & 3.1  & 1,3,14,16 \\
1997do & 3140 & 50766.21(0.45) &  -6 $\sim$ -4,7 $\sim$ 14   &  14.55(0.03) & 33.601(0.118) & 0.063 & 0.312(0.102) & 3.1  & 2,3,14 \\
1998V  & 5148 & 50591.27(0.84) &  -1 $\sim$ 5,11 $\sim$ 14   &  15.88(0.03) & 34.395(0.102) & 0.196 & 0.209(0.115) & 3.1  & 2,3,14 \\
1998ab & 8354 & 50914.43(0.25) &  -6                         &  16.08(0.05) & 35.213(0.100) & 0.017 & 0.394(0.082) & 3.1  & 2,3,14 \\
1998aq & 1354 & 50930.80(0.13) &  -2 $\sim$ 9                &  12.36(0.03) & 31.974(0.046) & 0.014 & 0.024(0.019) & 3.1  & 2,3,17 \\
1998bp & 3048 & 50936.39(0.33) &  -4 $\sim$ 3,12 $\sim$ 14   &  15.63(0.02) & 33.304(0.075) & 0.076 & 0.188(0.100) & 3.1  & 2,3,14 \\
1998de & 4671 & 51026.69(0.17) &  -6 $\sim$ 5                &  17.54(0.03) & 34.400(0.082) & 0.057 & 0.398(0.101) & 3.1  & 2,3,14,18 \\
1998dh & 2307 & 51029.83(0.22) &  -6 $\sim$ 2                &  14.15(0.03) & 32.846(0.087) & 0.068 & 0.471(0.061) & 2.76 & 2,3,14 \\
1998eg & 7056 & 51110.69(1.23) &  -2 $\sim$ 8                &  16.62(0.02) & 35.353(0.120) & 0.123 & 0.224(0.118) & 3.1  & 2,3,14 \\
1998es & 2868 & 51141.89(0.15) &  -4 $\sim$ 5,14             &  13.95(0.02) & 33.263(0.078) & 0.032 & 0.227(0.073) & 3.1  & 2,3,14 \\
1999aa & 4572 & 51231.97(0.15) &-6$\sim$3,7$\sim$11,13,14    &  14.86(0.02) & 34.468(0.042) & 0.040 & 0.020(0.018) & 3.1  & 2,3,7,14,19 \\
1999ac & 2943 & 51250.60(0.17) &  -6 $\sim$ 5,7 $\sim$ 14    &  14.27(0.02) & 33.334(0.081) & 0.046 & 0.308(0.072) & 3.1  & 2,3,14 \\
1999by & 827  & 51309.50(0.14) &  -6 $\sim$ -1,1 $\sim$ 10   &  13.66(0.02) & 31.158(0.073) & 0.016 & 0.170(0.083) & 3.1  & 2,3,20 \\
1999cc & 9452 & 51315.62(0.46) &  -5 $\sim$ 4                &  16.82(0.02) & 35.859(0.092) & 0.023 & 0.148(0.088) & 3.1  & 2,3,14,21 \\
1999dq & 4060 & 51435.70(0.15) &  -4 $\sim$ 0                &  14.85(0.02) & 33.668(0.067) & 0.110 & 0.369(0.077) & 3.1  & 2,3,14 \\
1999ej & 3831 & 51482.37(0.85) &  -3 $\sim$ 14               &  15.55(0.05) & 34.459(0.123) & 0.071 & 0.166(0.099) & 3.1  & 2,3,14 \\
1999gh & 2637 & 51513.53(0.91) &  3 $\sim$ 14                &  14.47(0.10) & 32.761(0.070) & 0.058 & 0.084(0.074) & 3.1  & 2,3,14 \\
1999gp & 7806 & 51550.30(0.17) &  -6 $\sim$ -3,1 $\sim$ 7    &  16.14(0.02) & 35.622(0.061) & 0.056 & 0.076(0.051) & 3.1  & 2,3,14,22 \\
2000cf & 10930& 51672.33(0.83) &  2 $\sim$ 7,13,14           &  17.18(0.04) & 36.382(0.104) & 0.032 & 0.194(0.097) & 3.1  & 2,3,14,21 \\
2000E  & 1266 & 51576.80(0.35) &  -6 $\sim$ -1               &  14.28(0.04) & 31.721(0.107) & 0.364 & 0.609(0.189) & 3.01 & 1,3,23,24 \\
2000cn & 6958 & 51707.83(0.16) &  -6,-5,7 $\sim$ 14          &  16.82(0.03) & 35.152(0.086) & 0.057 & 0.171(0.109) & 3.1  & 2,3,14 \\
2000dk & 4931 & 51812.47(0.28) &  -6 $\sim$ 6,8 $\sim$ 12    &  15.63(0.02) & 34.408(0.077) & 0.070 & 0.033(0.032) & 3.1  & 2,3,14 \\
2000fa & 6533 & 51892.25(0.57) &  1 $\sim$ 14                &  16.10(0.08) & 35.121(0.128) & 0.069 & 0.283(0.105) & 3.1  & 2,3,14 \\
2001V  & 4846 & 51973.28(0.16) &  -6 $\sim$ -3,4 $\sim$ 13   &  14.64(0.03) & 34.187(0.065) & 0.020 & 0.092(0.052) & 3.1  & 2,3,25 \\
2001el & 1102 & 52182.55(0.20) &  -6 $\sim$ 3,7 $\sim$ 11    &  12.85(0.02) & 31.544(0.077) & 0.014 & 0.696(0.059) & 2.4  & 1,3,26 \\
2002dj & 3131 & 52450.00(0.70) &  7 $\sim$ 14                &  14.30(0.04) & 33.315(0.185) & 0.096 & 0.384(0.186) & 3.1  & 1,27 \\
2002er & 2564 & 52524.84(0.15) &  -6 $\sim$ -3,1 $\sim$ 11   &  14.87(0.03) & 32.981(0.085) & 0.157 & 0.470(0.112) & 3.1  & 1,3,28,29 \\
2003du & 1994 & 52765.62(0.50) &  -6 $\sim$ 14               &  13.49(0.02) & 33.189(0.049) & 0.010 & 0.037(0.026) & 3.1  & 1,3,30 \\
2004eo & 4421 & 53279.20(0.50) &  -5 $\sim$ 14               &  15.51(0.03) & 34.188(0.124) & 0.109 & 0.000(0.082) $\dagger$ & 3.1  & 1,31 \\
2004S  & 2955 & 53039.87(0.25) &  6 $\sim$ 14                &  14.50(0.03) & 33.305(0.186) & 0.101 & 0.000(0.030) & 3.1  & 1,32 \\
2005cf & 2112 & 53534.00(0.30) &  -6 $\sim$ -1,2 $\sim$ 14   &  13.53(0.02) & 32.678(0.247) & 0.097 & 0.000(0.082) $\dagger$ & 3.1  & 1,33,34 \\
2005hk & 3548 & 53685.10(0.50) &-6$\sim$-1,2$\sim$6,11$\sim$14&  15.91(0.02) & 33.878(0.143) & 0.022 & 0.279(0.062) & 3.1  & 1,35 \\
2006gz & 6981 & 54020.20(0.50) &  -6 $\sim$ -3,4 $\sim$ 13   &  16.05(0.02) & 35.195(0.079) & 0.063 & 0.558(0.155) & 3.1  & 1,36 
\enddata
 
\tablenotetext{a}{$\text{m}^{max}_{B}$ is the apparent B maximum.}
\tablenotetext{b}{Milky Way dust extinction, from \cite{1998ApJ...500..525S}.}
\tablenotetext{\dagger}{Uncertain of $A_{V}(host)$ is set to $0.082$ which is the average value for the train supernovas.}
\tablerefs{(1) http://bruford.nhn.ou.edu/~suspect/index1.html;
(2) \cite{2008AJ....135.1598M};
(3) \cite{2007ApJ...659..122J};
(4) \cite{1983PASP...95...72B};
(5) \cite{1982SvAL....8..115T};
(6) \cite{1983ApJ...270..123B};
(7) \cite{2004MNRAS.349.1344A};
(8) \cite{1993ApJ...415..589K};
(9) \cite{1996MNRAS.278..111P};
(10) \cite{1995AJ....109.2121R};
(11) \cite{1996MNRAS.281..263M};
(12) \cite{1999AJ....117..707R};
(13) \cite{2001MNRAS.321..254S};
(14) \cite{2006AJ....131..527J};
(15) \cite{1999AJ....117.2709L};
(16) \cite{1998AJ....116.2431T};
(17) \cite{2005ApJ...627..579R};
(18) \cite{2001PASP..113..308M};
(19) \cite{2000ApJ...539..658K};
(20) \cite{2004ApJ...613.1120G};
(21) \cite{2006AJ....131.1639K};
(22) \cite{2001AJ....122.1616K};
(23) \cite{2003ApJ...595..779V};
(24) \cite{2001A&A...372..824V};
(25) \cite{2006PZ.....26....3T};
(26) \cite{2003AJ....125..166K};
(27) \cite{2008MNRAS.388..971P};
(28) \cite{2004MNRAS.355..178P};
(29) \cite{2005A&A...436.1021K};
(30) \cite{2005ApJ...632..450L};
(31) \cite{2007MNRAS.377.1531P};
(32) \cite{2007AJ....133...58K};
(33) \cite{2007MNRAS.376.1301P};
(34) \cite{2007A&A...471..527G};
(35) \cite{2007PASP..119..360P};
(36) \cite{2007ApJ...669L..17H}.
}
\label{tbl1}
\end{deluxetable}

\begin{deluxetable}{lllllllll}
\tabletypesize{\scriptsize}
\rotate
\tablecaption{Supernova and Host Galaxy Information.}
\tablewidth{0pt}
\tablehead{
\colhead{SN Ia}   &
\colhead{$cz_{cmb}$}   &
\colhead{$HJD_{B_{max}}$}   &
\colhead{$\text{m}^{max}_{B}$ \tablenotemark{a}}   &
\colhead{$\mu$}  &
\colhead{$E(B-V)_{gal}$\tablenotemark{b}}   &
\colhead{$A_{V,host}$}  &
\colhead{$R_{V,host}$}  &
\colhead{Ref} \\
& km/s & -240000(day) & \,\,\,\,\, mag & \,\,\,\,\,\,\,\,\,\, mag & \,\,\,\,\, mag & \,\,\,\,\,\,\,\, mag &  &
}
\startdata
1986G  & 803  & 46561.36(0.24) &  12.40(0.03) & 27.460(0.203) & 0.115 & 2.227(0.224) & 2.87 & 1,3,5,6  \\   
1989B  & 1051 & 47564.32(0.59) &  12.30(0.04) & 30.040(0.144) & 0.032 & 1.330(0.144) & 2.86 & 1,3,7,8  \\
1997dt & 1828 & 50785.60(0.31) &  15.62(0.05) & 32.723(0.191) & 0.057 & 1.849(0.198) & 3.03 & 2,3,4 \\
1998bu & 1204 & 50952.40(0.23) &  12.21(0.02) & 30.283(0.117) & 0.025 & 1.055(0.114) & 3.13 & 2,3,9,10\\
1998dk & 3609 & 51056.58(1.48) &  \dots \dots & 33.802(0.219) & 0.044 & 0.508(0.152) & 3.1  & 2,3,4 \\
1998dm & 1668 & 51062.03(1.16) &  \dots \dots & 33.067(0.178) & 0.044 & 1.045(0.160) & 3.08 & 2,3,4 \\
1998ec & 6032 & 51088.41(1.07) &  \dots \dots & 35.082(0.161) & 0.085 & 0.569(0.125) & 3.01 & 2,3,4 \\
1999cl & 2605 & 51342.28(0.26) &  15.03(0.04) & 30.604(0.163) & 0.038 & 2.666(0.160) &2.222 & 2,3,4,11 \\
1999gd & 5775 & 51518.40(1.14) &  17.02(0.03) & 34.606(0.167) & 0.041 & 1.381(0.156) & 2.99 & 2,3,4 \\
2000B  & 5977 & 51562.85(1.30) &  \dots \dots & 34.628(0.216) & 0.068 & 0.286(0.136) & 3.1  & 2,3,4 \\
2002bo & 1609 & 52356.89(0.14) &  14.05(0.05) & 31.945(0.105) & 0.025 & 1.212(0.099) & 2.57 & 1,3,12,13,14 \\
2003cg & 1591 & 52729.40(0.07) &  15.96(0.02) & 31.337(0.461) & 0.031 & 2.394(0.198) & 1.8  & 1,15       
\enddata
 
\tablenotetext{a}{$\text{m}^{max}_{B}$ is the apparent B maximum.}
\tablenotetext{b}{Milky Way dust extinction, from \cite{1998ApJ...500..525S}.}
\tablerefs{(1) http://bruford.nhn.ou.edu/~suspect/index1.html;
(2) \cite{2008AJ....135.1598M};
(3) \cite{2007ApJ...659..122J};
(4) \cite{2006AJ....131..527J};
(5) \cite{1987PASP...99..592P};
(6) \cite{1992A&A...259...63C};
(7) \cite{1994AJ....108.2233W};
(8) \cite{1990A&A...237...79B};
(9) \cite{1999AJ....117.1175S};
(10) \cite{1999IAUC.7298....2J};
(11) \cite{2006AJ....131.1639K};
(12) \cite{2004AJ....128.3034K};
(13) \cite{2003A&A...408..915S};
(14) \cite{2004MNRAS.348..261B};
(15) \cite{2006MNRAS.369.1880E}.
}
\label{tbl1x}
\end{deluxetable}

\subsection{Luminosity-flux ratio relations with one indicator}
\begin{figure}[htbp]
\begin{center}
\includegraphics[width=0.9\textwidth]{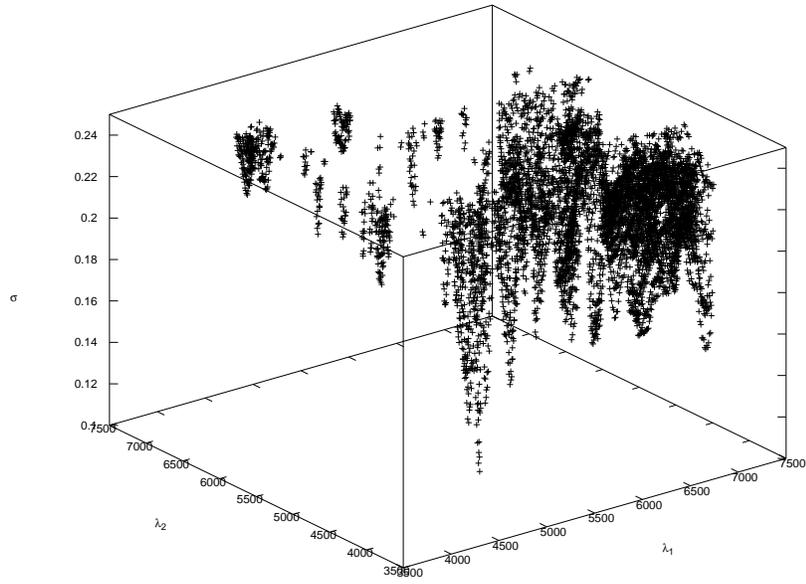}
\end{center}
\caption{Distribution of the magnitude scatter on the $(\lambda_1,\lambda_2)$
plane with $\sigma < 0.25$ at the phase $t=0$.}
\label{fig:trough}
\end{figure}

\begin{figure}[htbp]
\begin{center}
\includegraphics[width=0.9\textwidth]{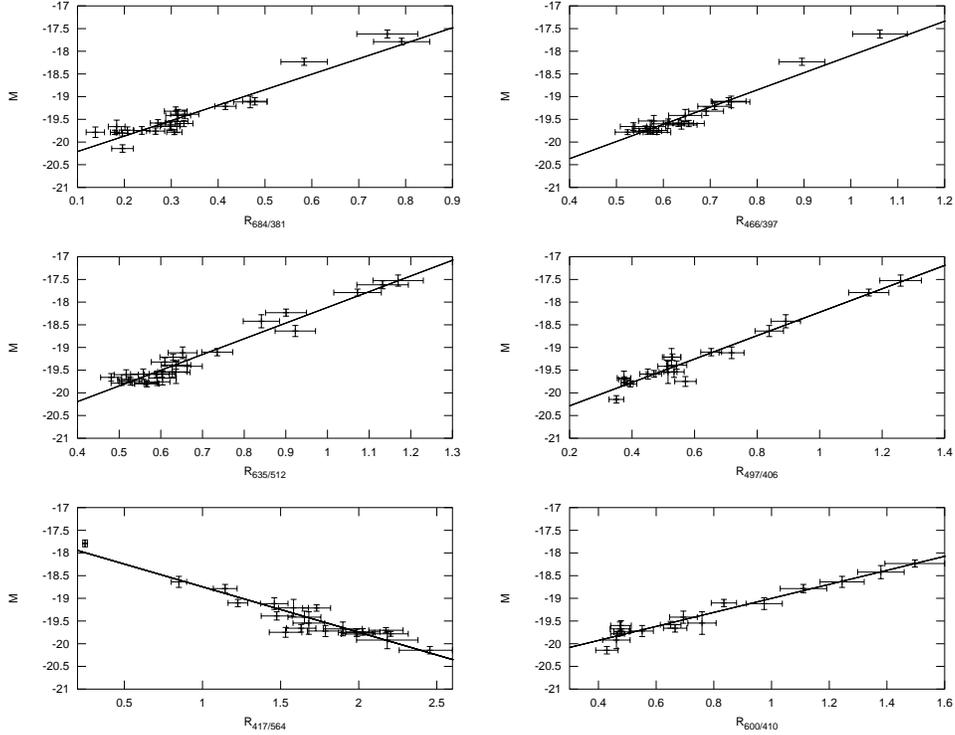}
\end{center}
\caption{$M \sim R_{\lambda_1/\lambda_2}$ diagrams with the local minimum scatter. From left to right
and top to bottom, the phases of the L-F relations plotted above are $(-3,0,3,6,9,12)$ respectively.}
\label{fig1}
\end{figure}

In this section, we investigate the relations between the B band maximum
 magnitude M and the flux ratio at each day within [-6,14] (here
 and after, when phase is mentioned it means that the spectra used
 are within $\pm 2.5$ days around these phases). At each phase t, we
 first select the supernovae spectra whose phase is within [t-2.5,t+2.5].
If there are multiple spectra for the same supernova within $\pm 2.5$ day phase window, we choose
the one which is nearest to the center phase. The wavelength range is [375, 750] nm.
It is noted that not every spectrum selected cover wavelength range [375, 750] nm,
for each wavelength pair, only the supernovas which cover these wavelengths are used.
Then we fit the data with linear relation
\begin{eqnarray}
\label{eq2}
M=sR_{\lambda_1/\lambda_2}+c,
\end{eqnarray}
for all possible wave pairs $(\lambda_1,\lambda_2)$ by the unweighted
least square method, and calculate the corresponding the scatter.
The number of the supernovas used in each fitting is required to be greater
than $18$ except $t=12$ and the flux ratio is defined by
$R_{\lambda_1/\lambda_2} = F(\lambda_1) / F (\lambda_2)$.
It is easy to find that the distribution of luminosity scatter in the $(\lambda_1,\lambda_2)$ plane
have some peaks and troughs. Figure \ref{fig:trough} is such an example at phase zero.
We search these troughs and find the minimum values. At
these places the mean values and uncertainties of $(s,c,\sigma)$ are calculated
through the following Probability Distribution Function (PDF)
\begin{eqnarray}
\label{sigma1}
{\cal L} (s,c,\sigma) \propto \prod_{i=1 \ldots n} \frac{1}{\sqrt{2 \pi \sigma^2}} \exp[-\frac{(M^i-sR^i_{\lambda_1/\lambda_2}-c)^2}{2 \sigma^2}],
\end{eqnarray}
where n is the number of the supernovas used in each calculation, i denotes the ith supernova.
In this process, two kind of methods are adopted, one is the unweighted
method, for which the errors of M or R are not accounted (in this case $\sigma$ at both 
hands of Equation \ref{sigma1} is the total scatter which contains the intrinsic scatter
and other uncertainties), the other is the weighted method for which the error of the magnitude, extinction,
distance modulus and spectrum flux are considered (in this case, $\sigma^2$ at right hand of Equation \ref{sigma1}
should be replaced by $\sigma^2_{t} = \sigma^2 + \sigma^2_{M} + s^2 \sigma^2_{R}$, $\sigma$
in $\sigma_{t}$ and ${\cal L} (s,c,\sigma)$ is the intrinsic scatter). The uncertain of the
flux ratio at some wave pairs contains two terms, one is that induced by the uncertain
of its host galaxy extinction, another is the noise and distortion of the
spectrum. It is noted that the train spectra are required to meet the condition
$|(B-V)^{syn} - (B-V)^{obs}| < 0.05$, so the relative error of the later term is estimated to be $5\%$.
Among these relations, five smaller ones at phases $t=-3,0,3,6,9,12$ are listed in Table \ref{tbl2}
(all of them are selected randomly besides the first one corresponds
to the best one in the unweighted case) and the first one are plotted in Figure \ref{fig1}.

\begin{figure}[htbp]
\begin{center}
\includegraphics[width=0.9\textwidth]{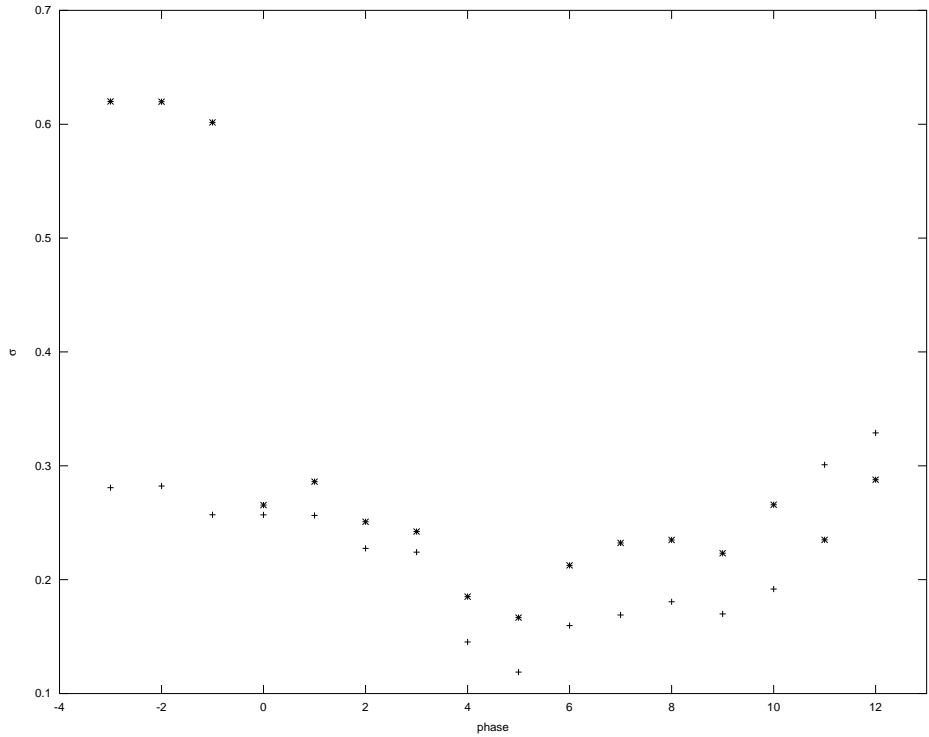}
\end{center}
\caption{Dependent of scatter (in the unweighted case) on the phase, the ``$\ast$'' points correspond to the $M \sim R_{500/409}$  relation, while the
``$+$'' points correspond to the $M \sim s1 R_{661/406} + R_{664/499} $ relation. Both above relations are local minimums at phase $t=5$.}
\label{vary}
\end{figure}

The results indicated that the $M \sim R_{\lambda_1/\lambda_2}$ relation exists not
only at the $t=0$ but also at other phases, for example in the unweighted case
the minimum scatters at $t=-3,3,6,9,12$ are $0.20(0.03)$ , $0.17(0.03)$, $0.17(0.03)$, 
$0.19(0.04)$, $0.14(0.03)$ respectively. At $t=0$, it is found that at the neighborhood of 
$(511,644)$, $(634,581)$ and $(638,408)$ the magnitude
scatter is small, which is consistent with the results of \cite{2009A&A...500L..17B} 
which use the spectra obtained by the Nearby Supernova Factory 
collaboration \cite{2002SPIE.4836...61A}, they find the small scatter at
$(642,512)$, $(577,642)$, $(642,443)$ and $(642,417)$. Besides, we also search 
the $M \sim log(R_{\lambda_1/\lambda_2})$ relation at each day within [-6,14]. Averagely
the scatter is slightly greater than the $M \sim R_{\lambda_1/\lambda_2}$ relation, for which
the similar conclusion is obtained in \cite{2009A&A...500L..17B}.  Last, how these L-F relations
dependent on the phase? Figure \ref{vary} illuminate two general cases, these
local minimum scatters at phase t=5 are smallest at t=5 and become larger at other phases. At
most case, one local minimum scatter found at one phase is not the local minimum
scatter at other phase. 

\begin{deluxetable}{lllll}
%\tabletypesize{\scriptsize}
\tabletypesize{\tiny}
%\rotate
\tablecaption{Five fitting results with smaller scatter at phases $t=-3,0,3,6,9,12$
using one flux ratio as the luminosity indicator. $(\bar{s}, \bar{c}, \bar{\sigma})$
is the results which do not consider the errors, while $(s, c, \sigma)$ considered
the errors. The number of supernovas used in each fitting is denoted by n.
All the results are listed in the form: mean value (1 $\sigma$ uncertainty).}
\tablewidth{0pt}
\tablehead{
\colhead{$phase (day)$}   &
\colhead{$(\lambda_1 / \lambda_2) (nm)$}   &
\colhead{$n$}   &
\colhead{$(\bar{s}, \bar{c}, \bar{\sigma})$} &
\colhead{$(s, c, \sigma)$} 
}
\startdata
-3&$684/381$& 23 &\{3.56(0.23),-20.58(0.09),0.20(0.03)\} & \{3.41(0.26),-20.55(0.09),0.14(0.04)\}\\
-3&$473/436$& 24 &\{2.69(0.23),-21.62(0.19),0.23(0.04)\} & \{2.70(0.21),-21.64(0.18),0.14(0.05)\}\\
-3&$641/438$& 24 &\{3.59(0.27),-21.05(0.13),0.23(0.04)\} & \{3.57(0.31),-21.04(0.14),0.15(0.05)\}\\
-3&$466/403$& 23 &\{3.51(0.30),-21.61(0.19),0.23(0.04)\} & \{3.54(0.34),-21.63(0.21),0.17(0.05)\}\\
-3&$557/438$& 24 &\{3.54(0.30),-21.65(0.20),0.24(0.04)\} & \{3.68(0.36),-21.72(0.22),0.14(0.06)\}\\
0&$466/397$& 19 &\{4.14(0.23),-22.10(0.15),0.13(0.03)\} & \{3.79(0.32),-21.88(0.20),0.05(0.04)\}\\
0&$497/430$& 20 &\{3.22(0.21),-21.93(0.17),0.16(0.03)\} & \{3.05(0.27),-21.80(0.21),0.07(0.05)\}\\
0&$588/431$& 20 &\{2.86(0.22),-21.47(0.16),0.18(0.03)\} & \{2.81(0.27),-21.45(0.19),0.11(0.05)\}\\
0&$634/399$& 19 &\{3.49(0.27),-20.76(0.11),0.18(0.04)\} & \{3.53(0.29),-20.77(0.11),0.08(0.04)\}\\
0&$633/431$& 20 &\{2.24(0.18),-20.81(0.12),0.18(0.04)\} & \{2.40(0.23),-20.91(0.14),0.10(0.04)\}\\
3&$635/512$& 26 &\{3.51(0.18),-21.60(0.13),0.17(0.03)\} & \{3.46(0.21),-21.57(0.14),0.06(0.04)\}\\
3&$687/443$& 26 &\{3.47(0.22),-20.43(0.09),0.21(0.03)\} & \{3.59(0.27),-20.47(0.10),0.15(0.04)\}\\
3&$660/582$& 26 &\{3.89(0.22),-21.90(0.16),0.21(0.03)\} & \{3.93(0.25),-21.93(0.17),0.08(0.05)\}\\
3&$660/404$& 25 &\{3.06(0.22),-20.31(0.09),0.22(0.04)\} & \{3.14(0.25),-20.33(0.10),0.17(0.04)\}\\
3&$531/403$& 25 &\{2.12(0.14),-20.41(0.09),0.22(0.03)\} & \{2.15(0.17),-20.43(0.10),0.18(0.04)\}\\
6&$497/406$& 20 &\{2.58(0.16),-20.80(0.10),0.17(0.03)\} & \{2.58(0.18),-20.80(0.11),0.10(0.05)\}\\
6&$476/442$& 21 &\{2.19(0.17),-21.34(0.17),0.22(0.04)\} & \{2.17(0.20),-21.33(0.19),0.18(0.05)\}\\
6&$594/446$& 21 &\{2.15(0.19),-20.87(0.15),0.25(0.04)\} & \{2.17(0.23),-20.87(0.17),0.20(0.05)\}\\
6&$633/446$& 21 &\{2.24(0.20),-20.70(0.14),0.25(0.04)\} & \{2.25(0.23),-20.70(0.15),0.21(0.05)\}\\
6&$634/509$& 21 &\{2.81(0.29),-21.41(0.23),0.28(0.05)\} & \{2.82(0.31),-21.41(0.23),0.22(0.06)\}\\
9&$417/564$& 21 &\{-0.99(0.09),-17.78(0.15),0.19(0.04)\} & \{-1.00(0.09),-17.75(0.16),0.15(0.05)\}\\
9&$635/407$& 21 &\{1.58(0.15),-20.29(0.09),0.20(0.04)\} & \{1.60(0.15),-20.30(0.08),0.14(0.04)\}\\
9&$499/389$& 21 &\{2.09(0.21),-20.48(0.12),0.22(0.04)\} & \{2.13(0.24),-20.50(0.13),0.19(0.05)\}\\
9&$525/425$& 21 &\{1.25(0.13),-20.69(0.14),0.23(0.04)\} & \{1.25(0.14),-20.69(0.15),0.19(0.05)\}\\
9&$639/444$& 21 &\{1.09(0.11),-20.26(0.10),0.22(0.04)\} & \{1.08(0.12),-20.25(0.10),0.20(0.04)\}\\
12&$600/410$& 17 &\{1.55(0.10),-20.55(0.09),0.14(0.03)\} & \{1.54(0.12),-20.54(0.09),0.07(0.05)\}\\
12&$498/433$& 17 &\{1.56(0.11),-20.89(0.11),0.15(0.03)\} & \{1.56(0.12),-20.88(0.11),0.06(0.04)\}\\
12&$489/416$& 17 &\{1.16(0.09),-20.49(0.10),0.16(0.04)\} & \{1.17(0.11),-20.50(0.11),0.13(0.05)\}\\
12&$597/515$& 17 &\{1.69(0.16),-20.97(0.16),0.19(0.04)\} & \{1.67(0.18),-20.95(0.18),0.15(0.05)\}\\
12&$638/451$& 17 &\{1.21(0.11),-20.25(0.09),0.19(0.04)\} & \{1.21(0.13),-20.26(0.11),0.17(0.05)\}
\enddata
 
%\tablenotetext{a}{}
\label{tbl20}

\end{deluxetable}

\subsection{Luminosity-flux ratio relations with two indicators}
\begin{figure}[htbp]
\begin{center}
\includegraphics[width=0.9\textwidth]{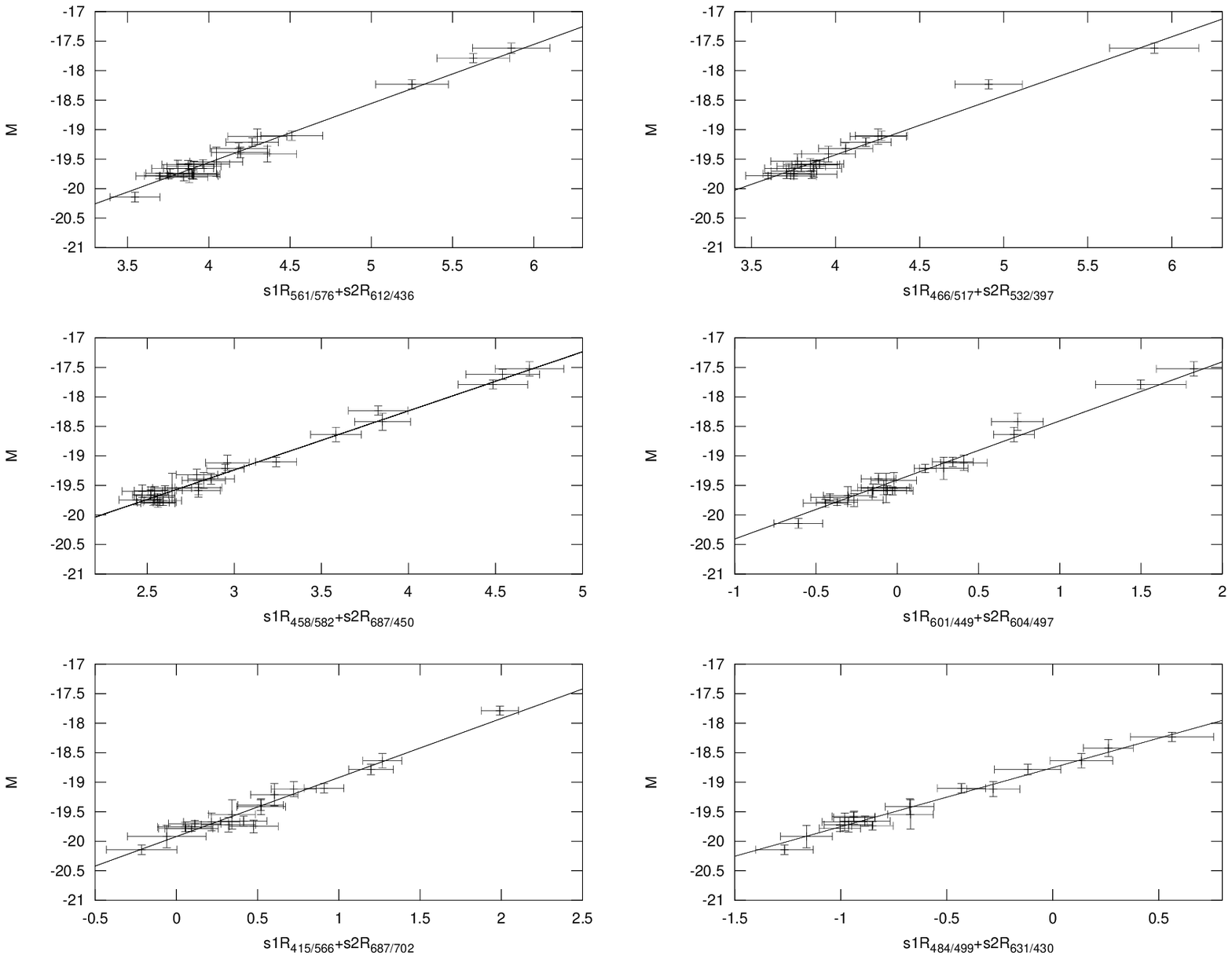}
\end{center}
\caption{$M \sim s1R_{\lambda_1/\lambda_2} + s2R_{\lambda_3/\lambda_4}$ diagrams
 with the local minimum scatter.  From left to right
and top to bottom, the phases of the L-F realtions plotted above are $(-3,0,3,6,9,12)$ respectively.}
\label{fig2}
\end{figure}

In this section we search the linear relations using two flux ratios as the luminosity
indicators. The same analyzing procedures are used as that in one luminosity indicator case,
the only difference is that two flux ratios are used in the fitting,
\begin{eqnarray}
\label{eq3}
M=s1R_{\lambda_1/\lambda_2}+s2R_{\lambda_3/\lambda_4}+c,
\end{eqnarray}
and the PDF of $(s1,s2,c,\sigma)$ is
\begin{eqnarray}
{\cal L} (s1,s2,,c,\sigma) \propto \prod_{i = 1 \ldots n} \frac{1}{\sqrt{2 \pi \sigma^2}} \exp[-\frac{(M^i - s1R^i_{\lambda_1/\lambda_2} - s2R^i_{\lambda_3/\lambda_4}-c)^2}{2 \sigma^2}].
\end{eqnarray}
The five better L-F relations are listed in Table \ref{tbl3} and the best one are plotted
in Figure \ref{fig2}. The minimum scatter at phases $t=-3,0,3,6,9,12$ are
respectively $0.10(0.02)$, $0.08(0.01)$, $0.10(0.02)$, $0.11(0.02)$,
$0.11(0.02)$, $0.09(0.02)$ in unweighted case. Compared with the results of one luminosity
indicator case all of them have large improvements: the magnitude
scatter decreases about $0.05 \sim 0.1$ mag. Again we test the linear relations
using two log flux ratios, the scatter is slightly greater than that using
the flux ratio directly as that in the one luminosity indicator case. Last, it is
noted that in both cases smaller scatters are obtained at phase 12 besides 0. 
\cite{2005ApJ...620L..87W} found that the supernova luminosity are better calibrated
by B - V color at t=12 days after B maximum. At some extent, color is the flux ratio
at the scale of broadband. Thus, this result may be not just a coincidence for
the train data set used here but a physical property for supernova.  

\begin{deluxetable}{lllll}
%\tabletypesize{\scriptsize}
\tabletypesize{\tiny}
%\rotate
\tablecaption{Five fitting results with smaller scatter for phases $t=-3,0,3,6,9,12$
using two flux ratios as the luminosity indicators. $(\bar{s1}, \bar{s2}, \bar{c}, \bar{\sigma})$
is the results which do not consider the errors, while $(s1, s2, c, \sigma)$ considered
the errors. The number of supernovas used in each fitting is denoted by n. All the results
are listed in the form: mean value (1 $\sigma$ uncertainty).}
\tablewidth{0pt}
\tablehead{
\colhead{$phase(day)$}   &
\colhead{$(\lambda_1 / \lambda_2, \lambda_3 / \lambda_4) (nm)$}   &
\colhead{$n$}   &
\colhead{$(\bar{s1}, \bar{s2}, \bar{c}, \bar{\sigma})$}  &
\colhead{$(s1, s2, c, \sigma)$}
}
\startdata
-3 & $561/576,612/436$ & 24 & \{2.75(0.11),3.84(0.33),-23.79(0.15),0.10(0.02)\} & \{2.58(0.22),3.75(0.64),-23.56(0.32),0.04(0.03)\} \\
-3 & $465/398,613/379$ & 23 & \{2.84(0.20),2.78(0.36),-22.22(0.12),0.11(0.02)\} & \{2.75(0.31),2.75(0.52),-22.15(0.17),0.04(0.03)\} \\
-3 & $474/435,503/398$ & 23 & \{2.25(0.14),2.56(0.32),-22.60(0.14),0.12(0.02)\} & \{2.27(0.22),2.31(0.46),-22.48(0.21),0.04(0.03)\} \\
-3 & $411/455,634/551$ & 23 & \{-1.76(0.13),3.21(0.21),-19.75(0.23),0.14(0.03)\} & \{-1.70(0.19),3.13(0.30),-19.76(0.36),0.05(0.04)\}\\
-3 & $433/561,546/576$ & 24 & \{-1.24(0.09),2.37(0.24),-19.93(0.34),0.15(0.03)\} & \{-1.22(0.13),2.28(0.29),-19.85(0.42),0.05(0.04)\}\\
0  & $466/517,532/397$ & 19 & \{3.06(0.25),4.33(0.15),-24.73(0.30),0.08(0.01)\} & \{1.95(0.47),4.29(0.47),-24.43(0.53),0.05(0.03)\} \\
0  & $475/438,591/519$ & 20 & \{3.11(0.12),2.73(0.28),-23.96(0.22),0.08(0.01)\} & \{2.99(0.31),2.32(0.60),-23.55(0.46),0.04(0.03)\}\\
0  & $461/518,699/439$ & 19 & \{1.92(0.23),4.93(0.20),-23.38(0.33),0.09(0.02)\} & \{1.44(0.36),4.82(0.57),-22.67(0.52),0.04(0.03)\} \\
0  & $473/519,634/439$ & 20 & \{1.76(0.26),2.91(0.14),-22.76(0.30),0.10(0.02)\} & \{1.32(0.38),2.82(0.23),-22.26(0.40),0.04(0.03)\} \\
0  & $466/397,674/582$ & 19 & \{3.08(0.35),1.18(0.34),-22.06(0.12),0.10(0.02)\} & \{2.81(0.45),1.35(0.49),-21.98(0.19),0.04(0.03)\} \\
3  & $458/582,687/450$ & 26 & \{0.83(0.07),5.32(0.18),-22.34(0.15),0.10(0.02)\} & \{0.78(0.12),5.31(0.38),-22.24(0.24),0.04(0.03)\} \\
3  & $461/514,700/441$ & 25 & \{1.71(0.21),2.33(0.18),-22.50(0.27),0.12(0.02)\} & \{1.48(0.28),2.46(0.28),-22.20(0.35),0.04(0.03)\} \\
3  & $578/462,690/395$ & 25 & \{-3.46(0.34),5.81(0.22),-18.74(0.17),0.12(0.02)\} & \{-3.17(0.42),5.68(0.34),-18.86(0.21),0.04(0.03)\} \\
3  & $461/576,701/407$ & 24 & \{0.72(0.09),4.26(0.17),-21.68(0.20),0.12(0.02)\} & \{0.68(0.10),4.28(0.25),-21.60(0.21),0.04(0.03)\} \\
3  & $574/401,637/404$ & 25 & \{-4.68(0.59),5.25(0.39),-19.25(0.14),0.11(0.02)\} & \{-3.02(0.66),4.21(0.42),-19.64(0.18),0.04(0.03)\} \\
6  & $601/449,604/497$ & 21 & \{3.44(0.13),-2.36(0.18),-19.40(0.12),0.11(0.02)\} & \{3.24(0.25),-2.21(0.25),-19.41(0.17),0.05(0.04)\} \\
6  & $498/387,604/384$ & 19 & \{3.39(0.24),-0.91(0.14),-20.78(0.07),0.11(0.02)\} & \{3.32(0.34),-0.91(0.21),-20.76(0.12),0.05(0.04)\} \\
6  & $497/390,605/429$ & 20 & \{3.63(0.24),-0.91(0.16),-20.67(0.07),0.12(0.02)\} & \{3.66(0.31),-0.90(0.22),-20.70(0.11),0.05(0.04)\} \\
6  & $574/419,589/389$ & 20 & \{-2.82(0.31),4.93(0.39),-20.51(0.07),0.14(0.03)\} & \{-2.40(0.49),4.47(0.54),-20.49(0.14),0.06(0.05)\} \\
6  & $604/497,634/451$ & 21 & \{-1.34(0.20),3.38(0.17),-19.97(0.15),0.14(0.03)\} & \{-1.23(0.22),3.27(0.24),-20.01(0.18),0.08(0.05)\} \\
9  & $415/566,687/702$ & 19 & \{-1.04(0.04),2.73(0.45),-20.87(0.52),0.11(0.02)\} & \{-1.02(0.08),1.88(0.59),-19.92(0.70),0.05(0.04)\}\\
9  & $421/556,685/617$ & 21 & \{-0.94(0.07),0.98(0.14),-18.96(0.19),0.12(0.02)\} & \{-0.92(0.08),1.03(0.15),-19.03(0.21),0.05(0.04)\} \\
9  & $609/471,660/706$ & 19 & \{-2.55(0.16),2.10(0.20),-24.43(0.38),0.12(0.02)\} & \{-2.50(0.34),1.75(0.25),-23.78(0.47),0.05(0.04)\} \\
9  & $422/518,684/620$ & 21 & \{-1.33(0.14),1.93(0.20),-19.34(0.27),0.13(0.02)\} & \{-1.33(0.18),1.82(0.26),-19.23(0.34),0.07(0.05)\}\\
9  & $565/406,576/499$ & 21 & \{1.61(0.10),-0.97(0.19),-19.54(0.17),0.14(0.03)\} & \{1.62(0.13),-0.96(0.20),-19.55(0.19),0.06(0.04)\} \\
12  & $484/499,631/430$ & 17 & \{-2.00(0.22),1.67(0.07),-18.48(0.24),0.09(0.02)\} & \{-1.74(0.43),1.67(0.16),-18.75(0.48),0.05(0.04)\} \\
12  & $605/421,610/525$ & 17 & \{2.75(0.19),-3.31(0.38),-19.97(0.05),0.10(0.02)\} & \{2.67(0.31),-3.18(0.67),-19.96(0.12),0.05(0.04)\} \\
12  & $541/431,634/499$ & 17 & \{1.27(0.07),-1.48(0.19),-19.64(0.14),0.10(0.02)\} & \{1.21(0.11),-1.36(0.30),-19.68(0.22),0.05(0.04)\} \\
12  & $500/431,547/616$ & 17 & \{1.64(0.09),0.27(0.04),-21.52(0.15),0.10(0.02)\} & \{1.63(0.14),0.27(0.06),-21.52(0.18),0.04(0.03)\} \\
12  & $599/417,622/499$ & 17 & \{1.25(0.07),-0.69(0.26),-19.86(0.19),0.12(0.03)\} & \{1.27(0.10),-0.70(0.28),-19.82(0.21),0.05(0.04)\}
\enddata

\label{tbl3}
 
%\tablenotetext{a}{}

\end{deluxetable}

It must to be stressed here that the definite position of troughs and their scatter may change
with more spectra are added into the train data and the coefficients and uncertainties of these L-F relations
are not very important in our method to constrain the distance and extinction of supernova, since they
do not depend on particular one of the L-F relations but many of them as you may see in the later section.

\section{Distance, host galaxy extinction of low redshift SN Ia}
\label{sn:fitting}
\subsection{Method}
\label{sn:method}
As we know in the above section, the absolute B band maximum magnitude M
 is $M=m-k-A_{B,gal}-A_{B,host}-\mu$. $A_{B,gal}$ can be calculated by
 $A_{B,gal} = E(B-V)_{gal} R_B$ where the Galaxy extinction $(E(B-V)_{gal},R_{B,gal})$
 is adopted from the \cite{1998ApJ...500..525S}. Besides this, $A_{B,host} = E(B-V)_{host} R_{B,host}$,
 so the right hand side of Eq. (\ref{eq1}) is a function of three unknown parameters
 $(E(B-V)_{host}, R_{B,host}, \mu)$. Since $R_B$ can be easily obtained from $R_V$ through
 the CCM extinction law \cite{1989ApJ...345..245C}, it is also a function of unknown
 parameters $(E(B-V)_{host}, R_{V,host}, \mu)$. At the same time we can also estimate the
 absolute B band maximum magnitude M from flux ratios. When ones use one flux ratio
 as the luminosity indicator, $M=sR_{\lambda_1/\lambda_2}+c$;
 while one uses two flux ratios as the luminosity indicators,
 $M=s1R_{\lambda_1/\lambda_2}+s2R_{\lambda_3/\lambda_4}+c$ (refer to section \ref{sn:relation}).
 Here the flux ratio is defined by $R_{\lambda_1/\lambda_2} = F(\lambda_1) / F(\lambda_2)$,
 with $F(\lambda)$ being the flux corrected for both Galaxy and host galaxy
 dust contamination at the SN rest frame, which can be computed from the observed
 flux $f(\lambda)$,
\begin{eqnarray}
F(\lambda) \propto f(\lambda (1+z)) 10^{(A_{\lambda (1+z),gal}+A_{\lambda,host})/2.5}.
\end{eqnarray}
Thus the flux ratio is
\begin{eqnarray}
R_{\lambda_1/\lambda_2} = \frac{f(\lambda_1 (1+z))}{f(\lambda_2 (1+z))}
10^{(A_{\lambda_1(1+z) ,gal}-A_{\lambda_2 (1+z),gal}+A_{\lambda_1,host}-A_{\lambda_2,host})/2.5},
\end{eqnarray}
where $A_{\lambda}$ is related to $A_V, R_V$ by $A_{\lambda}=A_V(a_{\lambda}+b_{\lambda}/R_V)$
 \cite{1989ApJ...345..245C} and $z$ is the redshift. So the right hand side of Eq. (\ref{eq2}) (or Eq. (\ref{eq3}))
 is a function of unknown parameters $(E(B-V),R_V)$ (here and in the following, $A,R,E$ refer
 to the host galaxy except stated otherwise.) when the Galaxy extinction is known.

With only one relation of Eq.(\ref{eq2}) or Eq. (\ref{eq3}) one can not determine
 the host galaxy extinction and distance modulus. But it should be noted that there
 exist many luminosity-flux ratio relations at the same supernova phase. Therefore,
 it is possible to constrain the host galaxy parameter and distance modulus when one
 applies many L-F relations (the distribution of the luminosity scatter $\sigma$ in the
 $(\lambda_1,\lambda_2)$ or $(\lambda_1,\lambda_2,\lambda_3,\lambda_4)$ plane has
 peaks and troughs, only the minimum of the troughs are used here. In later section,
it will be seen that when the number of L-F relations great than 20 the results 
will converge). Let us denote them as $(\lambda_1,\lambda_2)_{sn}$ in the one indicator case
 (or $(\lambda_1,\lambda_2,\lambda_3,\lambda_4)_{sn}$ in the two indicators case).
 Then the probability distribution function (PDF) of $(E(B-V),R_V,\mu)$ is
\begin{eqnarray}
&&{\cal L} (E(B-V),R_V,\mu) \propto \prod_{i \in (\lambda_1,\lambda_2)_{sn}}   \nonumber\\
&& \exp[-\frac{(s^iR^i_{\lambda_1/\lambda_2}+c^i-m+k+A_{B,gal}+A_{B,host}+\mu)^2}{2{\sigma^{i}}^2}]
\end{eqnarray}
in the case of one flux ratio as the luminosity indicator; while in the case of two flux ratios
as the luminosity indicators, the PDF is
\begin{eqnarray}
&&{\cal L} (E(B-V),R_V,\mu) \propto \prod_{i \in (\lambda_1,\lambda_2,\lambda_3,\lambda_4)_{sn}}  \nonumber\\
&& \exp[-\frac{(s1^iR^i_{\lambda_1/\lambda_2}+s2^iR^i_{\lambda_3/\lambda_4}+c^i-m+k+A_{B,gal}+A_{B,host}+\mu)^2}{2{\sigma^{i}}^2}].
\end{eqnarray}
Here $\sigma^{i}$ is the scatter obtained for the ith L-F relation.
It should be noted that the above formula is only right when $s,c,\sigma$ is fixed,
when they are distributed according to the function $f(s,c,\sigma)$, the PDF becomes
\begin{eqnarray}
&&{\cal L} (E(B-V),R_V,\mu) \propto \prod_{i \in (\lambda_1,\lambda_2)_{sn}} \int ds^i dc^i d\sigma^i  f(s^i,c^i,\sigma^i)  \nonumber\\
&&  \exp[-\frac{(s^iR^i_{\lambda_1/\lambda_2}+c^i-m+k+A_{B,gal}+A_{B,host}+\mu)^2}{2{\sigma^{i}}^2}],
\end{eqnarray}
or
\begin{eqnarray}
\label{eq4}
&&{\cal L} (E(B-V),R_V,\mu) \propto \prod_{i \in (\lambda_1,\lambda_2,\lambda_3,\lambda_4)_{sn}}   \int ds1^i ds2^i dc^i d\sigma^i  f(s1^i,s2^i,c^i,\sigma^i)  \nonumber\\
&& \exp[-\frac{(s1^iR^i_{\lambda_1/\lambda_2}+s2^iR^i_{\lambda_3/\lambda_4}+c^i-m+k+A_{B,gal}+A_{B,host}+\mu)^2}{2{\sigma^{i}}^2}].
\end{eqnarray}
From this, it is easy to calculate the mean values of $(E(B-V),R_V,\mu)$ and their uncertainties.
It should be stressed here that the above method is not only suitable for the L-F relation with two indicators
but also for the L-F relation with one indicator. However, at the same level of scatter, the number of first relations
are far less than the second ones. For the data used here, only the second case can give the real constraint on
distance and the host galaxy extinction. So in later section, we mainly focus on the L-F realtions with 
two indicators.
\subsection{Extinction parameter}
\label{sn:discussion2}
In this method, one of the extinction parameter $E(B-V)$ can be constrained very well, while $R_V$
can be only constrained loosely or even can not be constrained. It is noted that according to the
CCM extinction law
\begin{eqnarray}
A_{\lambda_1} - A_{\lambda_2} & = & A_V[(a_{\lambda_1}-a_{\lambda_2}) +(b_{\lambda_1}-b_{\lambda_2})/R_V]  \nonumber\\
& = & E(B-V) [R_V (a_{\lambda_1}-a_{\lambda_2}) + b_{\lambda_1}-b_{\lambda_2}],
\end{eqnarray}
In most cases $R_V |(a_{\lambda_1}-a_{\lambda_2})|$ is smaller than $|b_{\lambda_1}-b_{\lambda_2}|$.
 Thus, $(A_{\lambda_1} - A_{\lambda_2})$ and also the right hand side of Eq. (\ref{eq2}) or
 Eq. (\ref{eq3}) is mainly determined by $E(B-V)(b_{\lambda_1}-b_{\lambda_2})$. It also should be
 noted that $A_B \approx E(B-V) (R_V +1)$. Thus using a few relations Eq. (\ref{eq2}) or
 Eq. (\ref{eq3}), one can determine $E(B-V)$ very well, but cannot determine $R_V$ or only give
a weak constraint on it. Even it can give weak constraint on $R_V$, there exist degeneracy between $R_V$
 and $\mu$, since $(R_V,\mu)$ appear in the
 PDF in the form of $E(B-V)R_V + \mu$ approximately. In such a case, for a given $E(B-V)$, $(R_V,\mu)$
 can only be determined to a region along the line $E(B-V)R_V + \mu \approx const$. This 
can be seen from the contour map of $(R_V,M^*[=M+A_{B,gal}+A_{B,host}=m-k-\mu])$ for
SN 1998dm. However, $R_V$ also affects
 $(A_{\lambda_1} - A_{\lambda_2})$ and also the right hand side of Eq. (\ref{eq2}) or
 Eq. (\ref{eq3}). For different luminosity-flux ratio relations their importance are
 different. Thus $R_V$ still can be  determined from this method in principle, but
for the L-F relations obtained here, the constraint
on $R_{V}$ is weak and even not converge (approach to infinity or zero, since in the fitting
procedure $R_{V}$ is assumed to be great than zero). In order to break the degeneracy and give more better
 constraint on host galaxy extinction and distance, one needs more better
 luminosity-flux ratio relations which are derived from the SN Ia samples that the host galaxy
 extinction is better corrected. So in the fitting procedure, $R_V$ is always fixed at some
constant. This do not affect the constraint on the values of $E(B-V)$ largely.

\begin{figure}[htbp]
\begin{center}
\includegraphics[width=0.7\textwidth]{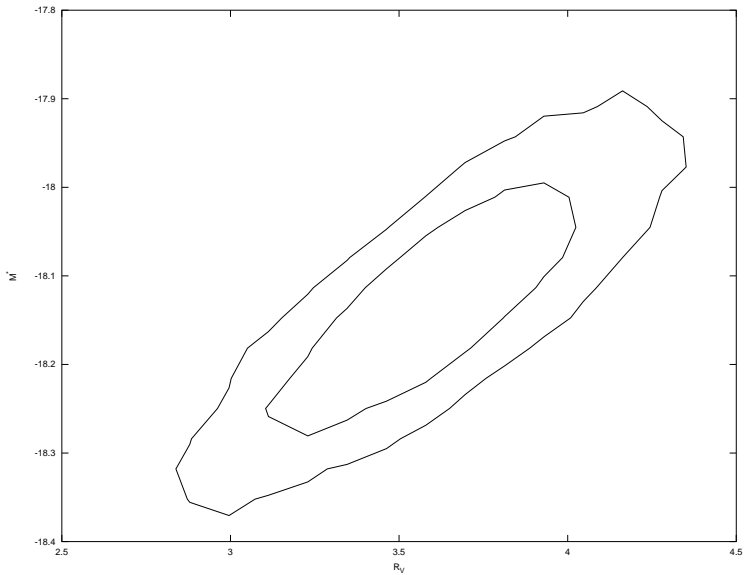}
\end{center}
\caption{Contour map of the effective magnitude $M^*$ of SN 1998dm and the 
extinction parameter $R_V$ of its host galaxy, the two regions represent $1 \sigma, 2 \sigma$
separately.}
\end{figure}

\subsection{Fitting procedures and results}
For a given spectrum with phase $t$, one can fit this spectrum with different
L-F relations whose phase is around $t$. Using one L-F realtions one get a result.
Among these results, one can choose the result which make the smallest change of the
spectrum (due to the host galaxy extinction) as the best one. 
The above choice means that in the case of large host galaxy extinction the result with
smallest $E(B-V)$ is the best one, while in the case of small extinction the result with
$E(B-V)$ that approaches zero is the best one. This result is always better than
the one obtained by applying the L-F relations at phase $t$. We also attempt
to choice the result which has the smallest $\chi^2 / dof$ as the best one, but it
does not work as well as the result which has the smaller change of the
spectrum. This point should be checked in a further study.

\begin{figure}[htbp]
\begin{center}
\includegraphics[width=0.7\textwidth]{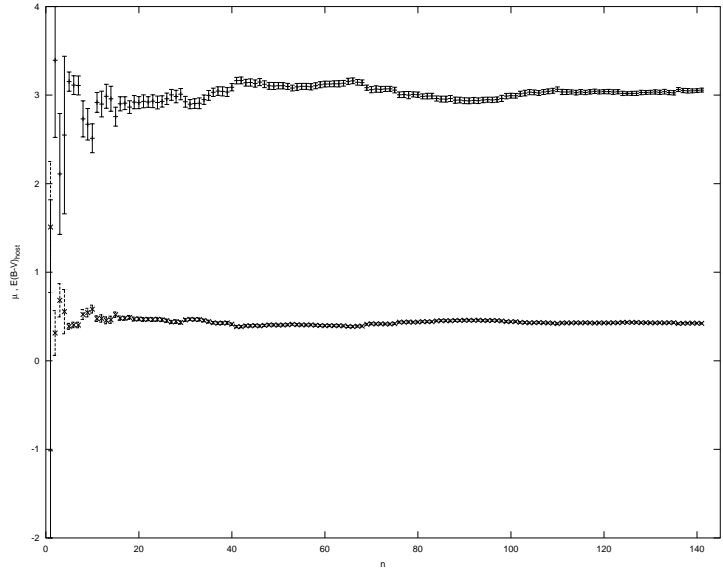}
\end{center}
\caption{Dependence of supernova distance and its host galaxy extinction of
SN 97dt on the number of L-F relations used in the fitting. The upper points
represent $(\mu-30,n)$, while the bottom points denote $(E(B-V),n)$.}
\label{dtx}
\end{figure}

Since the constraint on $R_V$ is weak, to fit the low redshift supernova spectra listed in Table \ref{tbl1x},
$R_{V}$ takes values provided there. Thus the fitting
parameter is $(E(B-V), \mu)$. When the apparent magnitude is not
known, we choose the fitting parameter as $(E(B-V), M^*=M+A_{B,gal}+A_{B,host}=m-k-\mu)$
(one can also choose $(E(B-V), M)$). Figure \ref{dtx} illuminate how $(E(B-V), \mu)$
converges with the number of L-F relations used in the fitting. 
Table \ref{res:lowz} present all the results of low supernova samples. 
Besides the best ones, the results determined by L-F relations
whose phase is given in Table \ref{tbl1x} are also listed. It is found that for most supernovas,
the first set of results is close to the values listed in Table \ref{tbl1x}, besides SN 1989B
whose host galaxy extinction is estimated to be larger than that estimated by \cite{2007ApJ...659..122J}
(correspondingly the distance is smaller than that estimated by \cite{2007ApJ...659..122J}).
Thus the method proposed in this section can indeed give information on the distance and its
host galaxy extinction.

It is noted that the distances obtained from the spectra with different phases
 are generally consistent with each other but not completely. On
 the one side, this fact reflects that the method is valid, on the other hand it
indicates there exist some uncertainty in the
 relations Eq. (\ref{eq2}) or Eq. (\ref{eq3}) which are obtained from a 38 SN
 Ia samples. This can be regarded as one advantage of the flux ratio method.
 That is, from the same method, different distances can be obtained to monitor
 the inconsistency of the method. The smaller is the difference between these
 distances, the more reliable to apply this method. Furthermore, one can get two
 different distances by using the same spectrum from luminosity-flux ratio
 Eq. (\ref{eq2}) and Eq. (\ref{eq3}) respectively, it also can reflect how well the F-L relations are.
In short, it is possible to see the reliability of the flux ratio method from itself
when one have two or more supernova spectrum with different phase.

\begin{deluxetable}{llllllll}
\tabletypesize{\scriptsize}
%\rotate
\tablecaption{Extinction and distance of low redshift supernovas.}
\tablewidth{0pt}
\tablehead{
\colhead{SN Ia}   &
\colhead{Date}  &
\colhead{$E(B-V)^1_{gal}$ \dag}   &
\colhead{$\mu^1$ or $M^{1*}$ \ddag}  &
\colhead{phase1 \$}  &
\colhead{$E(B-V)^0_{gal}$ \dag}   &
\colhead{$\mu^2$ or $M^{0*}$ \ddag}  &
\colhead{phase0 \$}
}
\startdata
1986G  & 1986-05-08 & 0.78(0.00) & 27.42(0.03) & -1& 0.84(0.01) & 27.12(0.04)  & -3\\
1989B  & 1989-02-17 & 0.64(0.01) & 29.12(0.05) & 9 & 0.94(0.01) & 28.19(0.08)  & 10\\
1997dt & 1997-12-04 & 0.42(0.01) & 33.06(0.05) & 1 & 0.42(0.01) & 33.06(0.05)  &  1\\
1998bu & 1998-05-28 & 0.27(0.01) & 30.60(0.03) & 9 & 0.27(0.01) & 30.60(0.03)  &  9 \\
1998bu & 1998-05-29 & 0.23(0.01) & 30.71(0.03) & 9 & 0.31(0.01) & 30.48(0.04)  & 10\\
1998dk & 1998-09-10 & 0.22(0.01) & 18.61(0.02) & 8 & 0.26(0.01) & 18.57(0.04)  & 10\\ 
1998dk & 1998-09-11 & 0.20(0.01) & 18.67(0.02) & 9 & 0.27(0.01) & 18.48(0.02)  & 11\\
1998dm & 1998-09-10 & 0.27(0.01) & 18.28(0.04) & 4 & 0.31(0.02) & 18.16(0.06)  & 5\\
1998dm & 1998-09-11 & 0.26(0.01) & 18.26(0.04) & 4 & 0.31(0.01) & 18.28(0.04)  & 6\\
1998ec & 1998-09-30 & 0.16(0.01) & 18.51(0.03) & 1 & 0.28(0.00) & 18.21(0.01)  & -1\\
1999cl & 1999-06-14 & 1.11(0.01) & 30.80(0.05) & 1 & 1.11(0.01) & 30.80(0.05)  & 1\\
1999gd & 1999-12-08 & 0.41(0.01) & 34.67(0.04) & 1 & 0.54(0.01) & 34.26(0.06)  &  4\\
2000B  & 2000-01-28 & 0.13(0.01) & 18.78(0.03) & 6 & 0.31(0.01) & 18.28(0.02)  & 9\\
2002bo & 2002-03-23 & 0.37(0.01) & 32.29(0.06) & 1 & 0.54(0.00) & 31.81(0.05)  & -1\\
2002bo & 2002-03-28 & 0.40(0.01) & 32.38(0.06) & 6 & 0.61(0.01) & 31.35(0.06)  & 5\\
2003cg & 2003-03-30 & 1.22(0.01) & 31.83(0.03) & 1 & 1.35(0.00) & 31.61(0.03) & -1
\enddata
 
\tablenotetext{\dag}{$E(B-V)^{1(0)}_{gal}$ are the host extinction obtained by using 
the L-F relation at phase1 (phase0).}
\tablenotetext{\ddag}{$\mu^{1(0)}$ are distance obtained by using the L-F relation
at phase1 (phase0).}
\tablenotetext{\$}{Phase0 is time of spectrum relative to the B magnitude maximum date
listed in Table \ref{tbl1}, while phase1 is the phase of the L-F relation , at which
the smallest change is needed to fit the spectrum.}

\label{res:lowz}
\end{deluxetable}

\section{Distance of high z SN Ia and Hubble diagram}
\label{sn:highz}

\subsection{Distance of high z SN Ia}
In this section, two high redshift supernova samples will be treated,
one of them is SCP spectra \cite{2005AJ....130.2788H}, the other is SNLS
three year spectra \cite{2009A&A...507...85B}. These samples are fitted with
the same method as the low redshift ones besides some difference due to their
high redshift. The observed high redshift spectra are blue in the rest frame than the low redshift
ones and their $S/N$ are smaller than the low redshift samples, especially
for the SNLS samples. So the L-F relations must be retreated again. 
First the spectrum flux of the train dataset are binned in
$c\varDelta \lambda / \lambda \sim 10000 km/s$ with $F(\lambda) =
\int_{\lambda - \Delta \lambda /2}^{\lambda + \Delta \lambda /2} 
f(\lambda) d\lambda / \varDelta \lambda$. Second, the L-F relations
are searched in the range of wavelength [375,550] nm. The first
point smoothes the features of the supernova spectrum, the second one
loses many possible L-F relations outside of wavelength range
[375,550] nm. Thus, generally speaking, the scatter of these new
L-F relations are averagely larger than the ones obtained above. The reason
to choice $c\varDelta \lambda / \lambda \sim 10000 km/s$ and
[375,550] nm is that the SNLS samples are smooth after such an average and their
wavelength in rest frame are mainly in the above range. The SCP samples
are better than the SNLS samples, but we still average them with 
$c\varDelta \lambda / \lambda \sim 10000 km/s$ and fit them with L-F
relations in the region [375,550] nm. 

\begin{figure}[htbp]
\begin{center}
\includegraphics[width=0.7\textwidth]{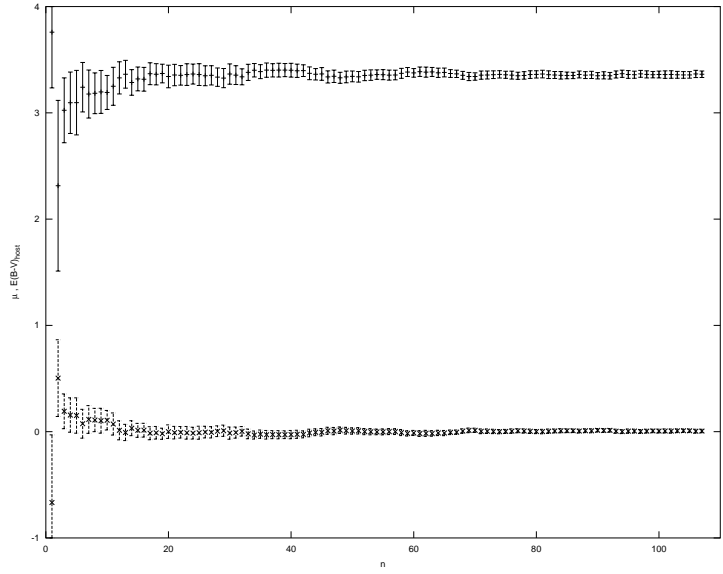}
\end{center}
\caption{Dependence of supernova distance and its host galaxy extinction of
SN 97ai on the number of L-F relations used in the fitting. The upper points
represent $(\mu-39,n)$, while the bottom points denote $(E(B-V),n)$.}
\label{aix}
\end{figure}

The host galaxy contamination in the high redshift supernova spectrum
 is more common than that in the low redshift one. For the SCP
samples, only the supernova spectrum which can be separated from the
host components are used here, they are 97am,97ai,97ac,97G,97aj, while
for the SNLS samples, we do not require this since the SNLS group play
many attentions in this respect. All the possible samples are required
in the preliminary phase [-3,12] which are determined in \cite{2005AJ....130.2788H}
and \cite{2009A&A...507...85B}, since out of this phase window, it is not sure whether
the spectrum can be fitted since the L-F relation is only searched in the
phase [-6,14]. Again, the supernova samples should has the public light
curves data to find the maximum light. After these conditions are considered, in
SNLS samples, only 03D1fl, 03D1gt, 03D4at, 03D4cx, 03D4cy,
03D4dy, 04D1ag, 04D1aj, 04D1ak, 04D2cf, 04D2fp, 04D2fs,
04D2gp, 04D2iu, 04D2ja, 04D4an, 04D4bk, 04D4bq, 04D4dw
left. Even the spectrum is averaged in $c\varDelta \lambda /
\lambda \sim 10000 km/s$, they are not smooth when $S/N < 1.0$. So the
spectra of 03D4cy, 04D1ak, 04D2ja, 04D4an are not used here due to their $S/N < 1.0$. The spectra of
03D1fl, 04D1aj, 04D4bk, 04D4dw are ignored ,because their wavelengths in the supernova 
rest frame are smaller than 500 nm, there are no enough better L-F relations
to get convergent results as stated before. Totally, there are 16 samples used 
in this section. These supernova spectra are processed by the same method
as that in the last section. The extinction and distance converges when
the number of L-F relations used in each fitting is greater than 20,
for example, Figure \ref{aix} illuminate how these two quantities of SN 97ai
vary with the number of L-F realtions used in the fitting. Last the
parameter $R_V$ takes a value of $2.2$ which is suggested by \cite{2009ApJS..185...32K}.
For the above high redshift supernova samples, $E(B-V)$ is small, so
different choice of $R_V$ only change the distance a little.

Here, two estimation of distance are adopted, the first is to estimate the
distance in the best fitting case, another is to average the few better
fitting results and the uncertain is half of the difference between the
largest result and the smallest result, for example, for 97am, there exists
several better fitting results, $[-0.06(0.01),41.89(0.02),7]$,
$[-0.02(0.02),41.99(0.02),8]$, $[-0.03(0.02),42.07(0.04),9]$,
$[-0.07(0.01),42.50(0.03),10]$ and $[-0.04(0.02),42.56(0.04),11]$ (
in order to better see the way to estimate the second distance, here the uncertain
of distance do not contains that from $\text{m}^{\text{max}}_{B}$; the results
are represented in the form [$E(B-V)$, $\mu$, phase of L-F relations used]). We choose
$[-0.02(0.02),41.99(0.02),8]$ as the best fitting results, while $42.20(0.33)$
as the second distance. It should be noted that since in every fitting, lots of L-F
relations are used, the uncertain due to it is very small, the main uncertain of distance
comes from the uncertain of $\text{m}^{\text{max}}_{B}$ and the difference of the 
results obtained by the L-F relations at different phase. For supernovas
which have only one better result, an extra uncertain of 0.20 mag is added into their distances.
This values is the average uncertain of distances for those supernovas which have
two or more better results. It is easy to see that the first
distance is overestimated while the second one is conservative. Both
of the two distances are given here, but only the extinction of the best fitting
results are given. It should be stressed that in the case of small extinction, the 
parameter $E(B-V)$ can not be regarded as the extinction like that in the case of large extinction,
but just an effective parameter. When the
extinction is small it contains many other effects besides the extinction.
Only in the large extinction case, it will dominate all the factors.

All of these spectra except 03D1gt are fitted very well. Too
large extinction are needed to fit the 03D1gt spectrum, which leads a very
small distance. The possible reason of this bad fitting will be discussed
later. The other supernova are fitted very well, they all do not need a large
extinction. Compared with the $\Omega_{m}=0.27,\Omega_{\Lambda}=0.73$ flat universe
the distance of $97aj$ has the largest deviation. The flat Universe predicts a
distance modulus of $42.84$, while the flux ratio method gives $\mu = 43.39(0.07)$.
However, \cite{2008ApJ...686..749K} estimates $\mu = 41.98(0.42)$  and
\cite{2009ApJ...700.1097H} takes $\mu = 42.067 (0.398)$, which deviate from $42.84$
far away than the value obtained by flux ratio method. Finally, we attempt to fit
the same supernovas by using the unweighted L-F relations, the results do not
change much.

\begin{deluxetable}{llllllllll}
\tabletypesize{\scriptsize}
\label{tbl2}
%\rotate
\tablecaption{Extinction and distance of high redshift supernovas.}
\tablewidth{0pt}
\tablehead{
\colhead{SN Ia}   &
\colhead{$z$}   &
\colhead{Date}   &
\colhead{$\text{m}^{\text{eff}}_{B}$ *}   &
\colhead{Phase0 \dag}   &
\colhead{$E(B-V)$ \ddag}  &
\colhead{$\mu^1$ \$}  &
\colhead{Phase1 \dag}  &
\colhead{$\mu^2$ \$}  &
\colhead{Phase2 \dag}
}
\startdata
97am  & 0.416 & 1997-03-13  & 22.52(0.04) & 9  & -0.02(0.02)  & 41.99(0.05) & 8 & 42.20(0.33) & 7 $\sim$ 11\\
97ai  & 0.450 & 1997-03-13  & 22.81(0.06) & 5  &  0.01(0.02)  & 42.36(0.07) & 7 & 42.49(0.29) & 6 $\sim$ 11\\
97ac  & 0.320 & 1997-03-14  & 21.83(0.04) & 11 &  0.01(0.02)  & 41.44(0.06) & 11& 41.42(0.06) & 10$\sim$ 14\\
97G   & 0.763 & 1997-01-13  & 24.49(0.41) & 4  & -0.02(0.01)  & 43.62(0.41) & 4 & 43.62(0.46) & 4 \\
97aj  & 0.581 & 1997-03-13  & 23.12(0.07) & -3 & -0.04(0.01)  & 42.39(0.07) & 2 & 42.48(0.11) & 2,3\\
03D4at& 0.633 & 2003-07-06  & 23.63(0.09) &5.5 & -0.10(0.02)  & 43.44(0.10) & 4 & 43.44(0.22) & 4\\
03D4cx& 0.949 & 2003-09-06  & 24.45(0.13) &2.7 &  0.00(0.02)  & 43.90(0.14) & 2 & 43.90(0.24) & 2\\
03D4dy& 0.604 & 2003-09-30  & 23.30(0.08) &4.8 &  0.00(0.02)  & 43.11(0.09) & 12& 43.04(0.28) & 11 $\sim$ 14\\
04D1ag& 0.557 & 2004-01-19  & 23.02(0.07) &4.3 & -0.02(0.03)  & 42.56(0.09) & 6 & 42.56(0.21) & 6\\
04D2cf& 0.369 & 2004-03-23  & 22.46(0.06) &8.5 &  0.01(0.03)  & 41.74(0.08) & 8 & 41.87(0.16) & 7 $\sim$ 10\\
04D2iu& 0.691 & 2004-05-12  & 24.26(0.13) &10.4& -0.05(0.02)  & 43.45(0.13) & 9 & 43.44(0.14) & 8,9\\
04D2fp& 0.415 & 2004-04-15  & 22.47(0.05) &1.8 & -0.02(0.01)  & 41.88(0.05) & -2& 41.97(0.30) & -4 $\sim$ 1\\
04D2fs& 0.357 & 2004-04-15  & 22.27(0.05) &1.7 &  0.05(0.01)  & 41.76(0.06) & 3 & 41.72(0.21) & 3\\
04D2gp& 0.707 & 2004-04-20  & 24.09(0.10) &2.7 &  0.00(0.02)  & 43.49(0.11) & 6 & 43.47(0.18) & 0 $\sim$ 6\\
04D4bq& 0.550 & 2004-07-16  & 23.40(0.08) &5.1 &  0.10(0.02)  & 42.68(0.09) & 6 & 42.62(0.31) & 4,5,6
\enddata

\tablenotetext{*}{$\text{m}^{\text{eff}}_{B}$=$m^{\text{max}}_{X}$ -$A_{X}$-$k_{XB}$, where X is some filter band. The first five
values adopted from \cite{1999ApJ...517..565P}, while others computed from the
first year SNLS data \cite{2006A&A...447...31A}.}
\tablenotetext{\dag}{Phase0 is time of spectrum relative to the B magnitude maximum date presented
in the reference \cite{2005AJ....130.2788H} or \cite{2009A&A...507...85B}, while phase1(phase2) is
the phase of the L-F relation, at which the smallest(smaller) change is needed to fit the spectrum.}
\tablenotetext{\ddag}{$E(B-V)$ is the effective host galaxy extinction obtained by using the L-F
relation at phase1.}
\tablenotetext{\$}{$\mu^{1(2)}$ are distance obtained by using the L-F relation
at phase1 (phase2).}

\end{deluxetable}

\subsection{Hubble diagram}
\label{sn:hubble}

\begin{figure}[htbp]
\begin{center}
\includegraphics[width=0.8\textwidth]{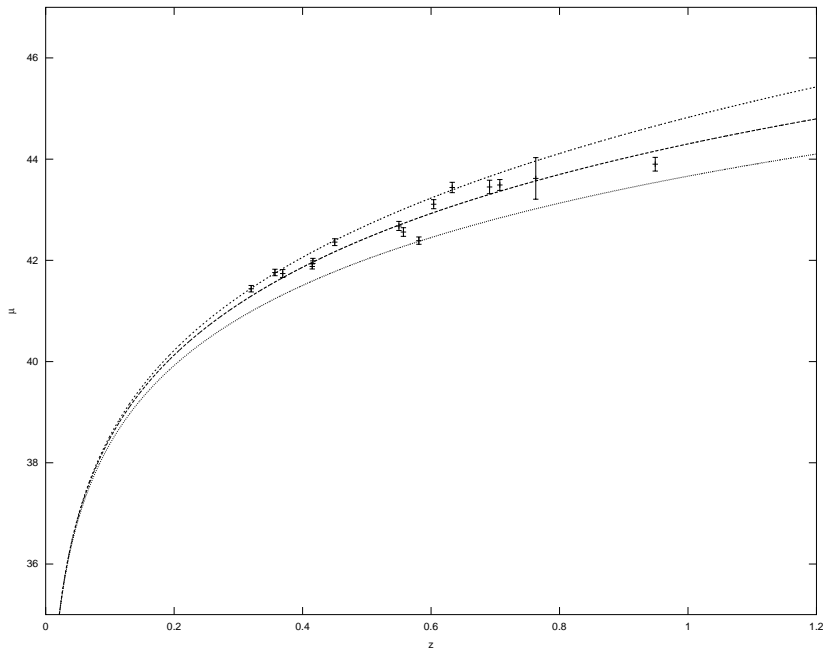}
\end{center}
\caption{Hubble diagram from the flux ratio method, the three lines
from top to bottom correspond to three cosmological models respectively: 1) flat
Universe with $\Omega_{m}=0,\Omega_{\Lambda}=1.0$; 2) flat Universe
with $\Omega_{m}=0.27,\Omega_{\Lambda}=0.73$; 3) flat Universe with
$\Omega_{m}=1.0,\Omega_{\Lambda}=0$.}
\label{Hubble}
\end{figure}

\begin{figure}[htbp]
\begin{center}
\includegraphics[width=0.8\textwidth]{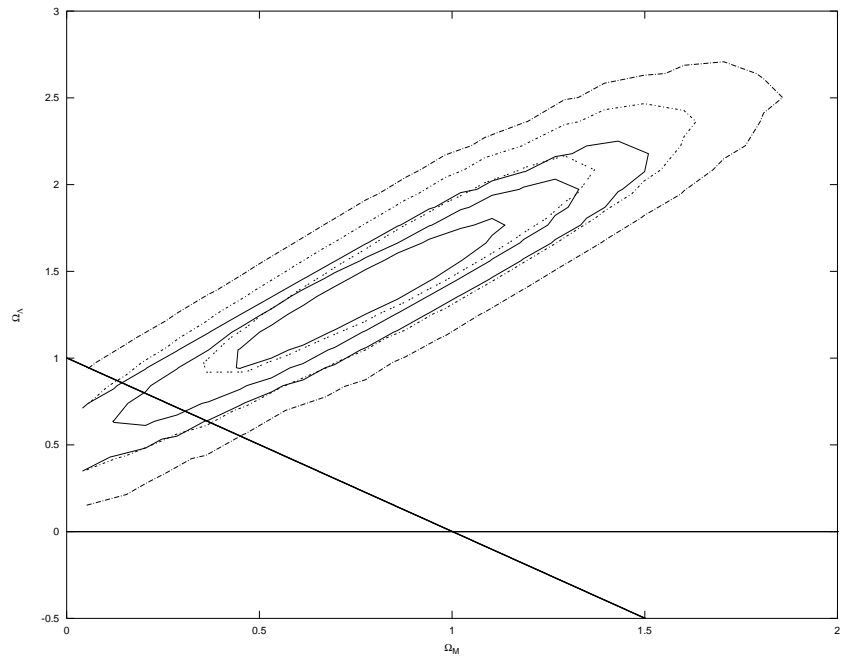}
\end{center}
\caption{$\Omega_{m},\Omega_{\Lambda}$ contours from the flux ratio method.
The solid/dashed lines represent $1 \sigma, 2 \sigma, 3 \sigma$ constraint
from two different distances respectively (one is overestimated while the other is conservative).}
\label{content}
\end{figure}

Using the distance obtained in the above section one can constrain the
contents of the Universe. The Hubble diagram and the contents of the
Universe are displayed in the Figure \ref{Hubble} and \ref{content}. Both of the above
two distances support an acceleration Universe. The contours
constrained from the two distances are consistent, but the contour from the first set of
distances is smaller than that from the second ones. The center of contour maps
for both cases are deviated from $\Omega_{m} + \Omega_{\Lambda} = 1$, but the
flat Universe is still in the $2 \sigma$ region. Considered that the number
of our high redshift supernovas is only 15 and the train spectra comes from
different observation, the flux ratio method can indeed be used to estimate the distance of
high redshift supernovas.

\section{Discussion and Summary}
\label{sn:discussion1}

From the above analysis, the distance of supernova and its host galaxy extinction
can be probed from the supernova spectrum, which is independent from other method.
Actually, most type Ia supernovas has spectroscopic follow-up, so it is possible to estimate their
distance and host galaxy extinction from the flux ratio method immediately. Furthermore,
when there are two or more spectra of one type Ia supernova, the results can be check with
each other. Another advantage of the method is that not one particular L-F relation but
many of them play the decisive role in the distance estimation.

Of course, there are a lot of possible factors which affect the flux ratio method. The most important
points are: 1) the supernova spectra train data set do not contain some one type of supernova spectrum, then
the L-F relations may not fit this type of supernova rightly; 2) the total distortion of the supernova
spectrum in the spectrograph, this can be monitored by checking the difference between the observed
magnitude and that synthesized from spectra;  3) the host galaxy contamination, lots of supernova
spectra have a large fraction of host galaxy component. To exploit these spectra, it should
subtract the host galaxy components from them cleanly. Any case mentioned above will leads
to wrong or bad results, may be the SNLS supernova 03D1gt is such a case.
The other possible factors to affect the method includes: 4) the redshift uncertain , both due to the
peculiar velocity and the measurement, since in the procedure of deredshift, the error of redshift will
leads some uncertain of the spectra in the rest frame; 5) extinction of Milky Way and host galaxy, the
host galaxy extinction for the train spectra will affect the L-F relations and then affect the distance
estimation of other supernova, the uncertain in Milky dust extinction also will leads to systematic error
in the distance estimation; 6) the noise of the spectrum in the measurement; 7) the uncertain of the phase of the
train spectra. All these factors will
affect the L-F relation, which make it easy to understand the following things. First, the host galaxy
extinction $E(B-V)$ is only an effective parameter which contains not only the effect of host galaxy
extinction but also other factors which may change the spectrum. When the host galaxy extinction is large,
it will dominate among all the possible factors as found in section \ref{sn:fitting}. Second, there is no
or only weak constraint on $R_V$. It is stated in the previous section that this method depends on $R_V$
loosely even in the ideal case. This dependence will be easily broken by the factors listed above, thus
only in a more precise analysis, the flux ratio method can provide an effective constraints on $R_V$.   

Finally, let us summary the contents of this paper. First, we generalized the flux ratio method
presented in \cite{2009A&A...500L..17B} to supernova spectrum phase which is not around the
date of B magnitude $t_{0}$. This is useful when one wants to use this method to study the
universe expansion, since generally speaking there are a lot of supernovas which do not have
spectrum around the phase $t_{0}$ but have spectrum at other phase. Second, we search the optimal
linear relationships between B maximum magnitude and flux ratios using two
flux ratios as the luminosity indicators and obtained a scatter which is small than 0.1. The absolute value of
the scatter is already small, but more important is that it will improve the scatter compared with
that obtained by using only one flux ratio as the luminosity indicator. This improvement possibly
comes from two sides, one is that in some case the intrinsic scatter in one flux ratio case can be
offset by introducing another flux ratio, the other is that the scatter due to the spectrum  which
is not at the same phase maybe be decreased in this method. In
\cite{2009A&A...500L..17B}, the linear relation using one flux ratio as the luminosity indicator
is studied at the phase $t=0$ on a better spectra samples obtained by the Nearby Supernova
Factory collaboration \cite{2002SPIE.4836...61A}, whose scatter can be smaller than $0.12$.
The result is already very good, but we believe that after using two flux ratios as luminosity
indicators, the scatter can be decreased further to a very low level. Third, we try to correct the host galaxy
extinction directly in this paper. \cite{2009A&A...500L..17B} try to use flux rations which corrected for
the host galaxy extinction and intrinsic dispersion without distinguishing the two effects. Here, we
have a general host galaxy dust correction, but introduce errors from the uncertainties associated with $A_V$.
Since the flux ratio is the ratio of flux at two wavelengths, it is less sensitive to the change
of the host galaxy extinction as the flux itself, but the host galaxy dust is still
an obstacle to standardize the supernova peak luminosity in the flux ratio method. Without
a better host galaxy extinction correction for the supernova samples, one can not get the precise
luminosity-flux relation relations. The best method to solve this problem is to redo the analysis in
a supernova samples whose dust extinction is small. Fourth, after searching the luminosity-flux
ratio relations, we attempt to use it to fit the distances and host galaxy extinctions of low redshift supernovas.
The results are consistent with that obtained by other independent
methods. However, there exists some degeneracy between distance
and the host galaxy extinction, which needs a further study. Last, the same method is applied
to the high z supernova samples. Before
fitting, we research the L-F relations using a smoother average condition in a small
wavelength range. Two kind of distances are estimated, from which
we obtained the Hubble diagram and analyzed the contents of the Universe, which support an acceleration Universe. 
So the flux ratio method can indeed provide another independent way to constrain
the distance of SN Ia and its host galaxy extinction.

\begin{acknowledgments}
We would like to thank Dr Wei-Min Sun and Shi Qi for improving the
manuscript. This research has made use of the CfA Supernova Archive
which is funded in part by the National Science Foundation through
grant AST 0606772 and the SNANA software which is developed to analyze
the SDSS supernova data. This research was supported by the Purple
Mountain Observatory Research Funds Y009C11054 and Natural Science Foundation
of China under Grant No. 10973039.
\end{acknowledgments}

%\bibliographystyle{apj}

%\bibliography{ref.bib}

\end{document}